\documentclass[10pt]{article}
\usepackage[english]{babel}
\usepackage{amsmath,amsfonts,amssymb,bm}\interdisplaylinepenalty=2500
\ifx\pdfoutput\undefined
  \usepackage{graphicx,color}   
  \usepackage[nativepdf,backref,bookmarks]{hyperref} 
  \usepackage{rotating}
\else
  \usepackage[backref,bookmarks]{hyperref}  
  \usepackage[pdftex]{graphicx,color}     
  \usepackage[pdftex]{rotating}              
\fi
\graphicspath{{figs/}}
\mathchardef\varphi="011E        \mathchardef\phi="0127
\ifx\undefined\degrees\def\degrees{\ensuremath{^{\circ}}}\fi
\ifx\undefined\celsius\def\celsius{\ensuremath{^{\circ}\mathrm{C}}}\fi
\ifx\undefined\unit\def\unit#1{\ensuremath{\mathrm{\,#1}}}\fi
\ifx\undefined\micro\def\micro{\ensuremath{\mu}}\fi
\ifx\undefined\sups\def\sups#1{\ensuremath{^{\mathrm{#1}}}}\fi
\ifx\undefined\subs\def\subs#1{\ensuremath{_{\mathrm{#1}}}}\fi
\ifx\undefined\ohm\def\ohm{\ensuremath{\mathrm{\Omega}}}\fi
\def\req#1{(\ref{#1})}

\begin{document}
\title{On the measurement of frequency and of its sample variance with high-resolution counters}
\author{Enrico Rubiola%
\thanks{Universit\'e Henri Poincar\'e, Nancy, France. e-mail~\texttt{enrico@rubiola.org}}
}
\date{Rev.\ 2.0,~~\today.~~arXiv physics/0411227}

\maketitle

\begin{abstract}
A frequency counter measures the input frequency $\overline{\nu}$ averaged 
over a suitable time $\tau$, versus the reference clock.  
High resolution is achieved by interpolating the clock signal.
Further increased resolution is obtained by averaging multiple frequency measurements highly overlapped.  In the presence of additive white noise or white phase noise, the square uncertainty improves from $\smash{\sigma^2_\nu\propto1/\tau^2}$ to $\smash{\sigma^2_\nu\propto1/\tau^3}$.
Surprisingly, when a file of contiguous data is fed into the formula of the two-sample (Allan) variance $\smash{\sigma^2_y(\tau)=\mathbb{E}\{\frac12(\overline{y}_{k+1}-\overline{y}_k)^2\}}$ of the fractional frequency fluctuation $y$, the result is the \emph{modified} Allan variance mod~$\sigma^2_y(\tau)$.  But if a sufficient number of contiguous measures are averaged in order to get a longer $\tau$ and the data are fed into the same formula, the results is the (non-modified) Allan variance.  Of course interpretation mistakes are around the corner if the counter internal process is not
well understood.  
\end{abstract}

\tableofcontents

\subsubsection*{Revision history}
\noindent\textbf{Rev.\ 1.0}, Nov 25, 2004. First draft.\\
\noindent\textbf{Rev.\ 2.0}, Dec 31, 2004.
Table 1 updated. Added Section 2.4 and Fig.\ 4.

\section{Background}\label{sec:background}
Let $v(t)=\sin2\pi\nu t=\sin[2\pi\nu_{00}t+\phi(t)]$ the input signal, where $\phi(t)$ is the phase fluctuation, $\nu_{00}=1/T_{00}$ is the nominal frequency (the double subscript `00', as in  $\nu_{00}$, is used to avoid confusion with the 0-th term of a time series; thus, $\nu_{00}$ is the same as $\nu_0$ commonly used in the literature, etc.), and
$\smash{\nu(t)=\nu_{00}+\frac{\dot{\phi}}{2\pi}}$ is the instantaneous frequency.  Finally, let $\smash{x(t)=\frac{\phi}{2\pi\nu_{00}}}$ the phase time fluctuation, i.e., the time jitter, and $\smash{y(t)=\frac{\dot{\phi}}{2\pi\nu_{00}}=\frac{\nu-\nu_{00}}{\nu_{00}}  }$ the fractional frequency fluctuation.  The notation used in this article is the same of the general references about frequency stability and noise \cite{barnes71im,rutman78pieee,ccir90rep580-3,ieee99std1139}. 

Denoting with $\mathbb{E}\{\cdot\}$ the expectation, the (classical) variance of $y$ is  $\sigma^2_y=\smash{\mathbb{E}\{[y-\mathbb{E}\{y\}]^2\}}$. In the presence of slow random phenomena, the variance depends on the measurement time and on the number of samples.  This is related to the fact that the algorithm used is a filter in the frequency domain, whose lower cutoff frequency is set by the number of samples.  Other variances are to be used, based on the idea that the estimator, clearly specified, has a lower cutoff frequency that blocks the dc and long-term components of noise.  There results a variance which is a function of the measurement time $\tau$.  Table~\ref{tab:variances} shows the spectral properties of the Allan variance and of the modified Allan variance, defined underneath. 

\newcommand{\pbx}[1]{\parbox{7ex}{\rule[0ex]{0ex}{2.5ex}#1\rule[-1.5ex]{0ex}{1.5ex}}}
\begin{table}[t]
\begin{center}
\caption{Noise types, power spectral densities, and Allan variances.%
  \vrule width 0pt height 1.5ex depth 1.5ex}\label{tab:variances}
\begin{tabular}[c]{lccccc}\hline\hline
\pbx{noise\\type}           &$S_\phi(f)$   &$S_y(f)$      &$S_\phi\leftrightarrow S_y$
	&$\sigma_y^2(\tau)$ & $\mathrm{mod}\,\sigma_y^2(\tau)$\rule[-1.5ex]{0ex}{4ex}\\\hline\hline
\pbx{white PM} &$b_0$         &$h_{2}f^{2}$  &$h_{2}=\frac{b_0}{\nu_0^2}$
	&\parbox{8ex}{\rule[0ex]{0ex}{0.5ex}
	    $\frac{3f_Hh_2}{(2\pi)^2}\:\tau^{-2}$\\
	    {\footnotesize $2\pi\tau f_H{\gg}1$}\rule[-0.5ex]{0ex}{1ex}} 
	&$\frac{3f_H\tau_0h_2}{(2\pi)^2}\:\tau^{-3}$\rule[-1.5ex]{0ex}{4ex}\\\hline
\pbx{flicker PM} &$b_{-1}f^{-1}$&$h_{1}f$      &$h_{1}=\frac{b_{-1}}{\nu_0^2}$
	&\parbox{14ex}{\rule[0ex]{0ex}{3ex}\centering
	     \hspace*{-1.5ex}\hbox{\footnotesize$[1.038+3\ln(2\pi f_H\tau)]$}\\
	     $\times\frac{h_1}{(2\pi)^2}\:\tau^{-2}$
	     \rule[-2ex]{0ex}{2ex}} 
	&\parbox{8ex}{$0.084\,h_1\:\tau^{-2}$\\$n{\gg}1$}\\\hline
\pbx{white FM} &$b_{-2}f^{-2}$&$h_0$         &$h_{0}=\frac{b_{-2}}{\nu_0^2}$
	&$\frac{1}{2}h_0\,\tau^{-1}$ 
	&$\frac{1}{4}h_0\,\tau^{-1}$\rule[-1.5ex]{0ex}{4ex}\\\hline
\pbx{flicker FM} &$b_{-3}f^{-3}$&$h_{-1}f^{-1}$&$h_{-1}=\frac{b_{-3}}{\nu_0^2}$
	&$2\ln(2)\;h_{-1}$ &$\frac{27}{20}\ln(2)\;h_{-1}$\rule[-1.5ex]{0ex}{4ex}\\\hline
\pbx{random\\walk FM} &$b_{-4}f^{-4}$&$h_{-2}f^{-2}$&$h_{-2}=\frac{b_{-4}}{\nu_0^2}$
	&$\frac{(2\pi)^2}{6}h_{-2}\tau$ 
	&$0.824\,\frac{(2\pi)^2}{6}h_{-2}\,\tau$\rule[-1.5ex]{0ex}{4ex}\\\hline
\pbx{\hbox{frequency drift $\dot{y}=D_y$}} &&&
	&$\frac12\,D_y^2\,\tau^2$ 
	&$\frac12\,D_y^2\,\tau^2$\rule[-1.5ex]{0ex}{4ex}\\\hline
\multicolumn{6}{l}{Here $\nu_{00}$ is replaced with $\nu_0$ for consistency with the general literature.}\\
\multicolumn{6}{l}{$f_H$ is the high cutoff frequency, needed for the noise power to be finite.}\\
\multicolumn{6}{l}{The columns $\sigma^2_y(\tau)$ and $\mathrm{mod}\,\sigma^2_y(\tau)$ are from \cite[p.\:79]{kroupa:frequency-stability} (adapted).}\\\hline\hline  
\end{tabular}
\end{center}
\end{table}

\subsection{Allan variance (AVAR)}
\begin{figure}[t]
\centering\includegraphics[scale=0.8]{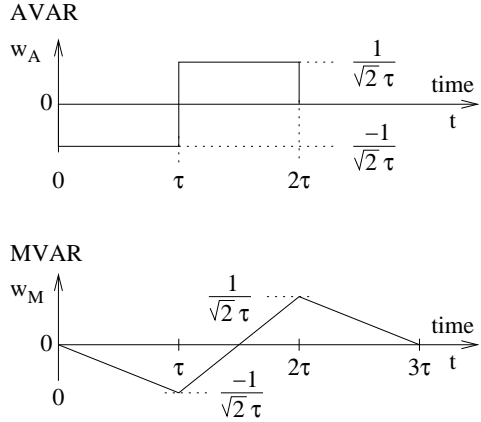}
\caption{Weight functions for the two-sample variance and for the modified Allan variance.}
\label{fig:wavelets}
\end{figure}

Originally, the Allan variance was introduced as a measurement tool for the frequency fluctuation of atomic clocks~\cite{allan66pieee}.  Given a stream of contiguous data $\overline{y}_k$ averaged on a time $\tau$, the simplest variance is the (classical) variance evaluated on two samples, $\smash{\sigma^2_y(\tau)=\frac12[y_{k+1}-y_k]^2}$.  The estimated variance is
\begin{align}
\sigma^2_y(\tau)
&=\mathbb{E}\left\{\frac{1}{2}\Bigl[\overline{y}_{k+1}-\overline{y}_{k}\Bigr]^2\right\}~,
\qquad\text{AVAR}
\label{eqn:avar-def}
\end{align}
and, expanding the time average, 
\begin{align}
\sigma^2_y(\tau)
&=\mathbb{E}\left\{\frac{1}{2}\biggl[
\frac{1}{\tau}\int_{(k+1)\tau}^{(k+2)\tau}y(t)\,dt - \frac{1}{\tau}\int_{k\tau}^{(k+1)\tau}y(t)\,dt
\biggr]^2\right\}~.
\label{eqn:avar-expanded}
\end{align}
The above can be rewritten as 
\begin{align}
\sigma^2_y(\tau)
&=\mathbb{E}\left\{\Bigl[\int_{-\infty}^{+\infty} y(t)\,w_A(t) \,dt \Bigr]^2\right\}
\label{eqn:avar-wavelet}\\
w_A&=\begin{cases}
     -\frac{1}{\sqrt{2}\tau} & 0<t<\tau\\
     \frac{1}{\sqrt{2}\tau} & \tau<t<2\tau\\
     0 & \text{elsewhere}
  \end{cases}
  \qquad\text{(Fig.~\ref{fig:wavelets})}
\label{eqn:avar-weight}
\end{align}
which is similar to a wavelet variance.  The weight function differs from the Haar wavelet in that it is normalized for power instead of energy. In fact, the energy of $w_A$ is
\begin{align}
\mathcal{E}\{w_A\}&=\int_{-\infty}^{+\infty} w^2_A(t)\,dt = \frac{1}{\tau}~.
\end{align}
while the energy of a wavelet is $\mathcal{E}\{\cdot\}=1$.

In the frequency domain, the AVAR is similar to a half-octave bandpass filter with the peak at the frequency of $\frac{1}{2\tau}$.

\subsection{Modified Allan variance (MVAR)}
Another type of variance commonly used in time and frequency metrology is the modified Allan variance $\smash{\mathrm{mod}\,\sigma^2_y(\tau)}$ \cite{allan81fcs,lesage84im}.  This variance was first introduced in the domain of optics~\cite{snyder80ao} because it divides white phase noise from flicker phase noise, which the AVAR does not. This is often useful in fast measurements.  MVAR is also related to the sampling theorem and to the aliasing phenomenon \cite{vernotte98metro-a,vernotte98metro-b} because the trigger samples the input process at a rate $1/\tau_0$ (see below).  The MVAR is defined as
\begin{align}
\begin{split}
&\mathrm{mod}\,\sigma^2_y(\tau)=\mathbb{E}\left\{\frac{1}{2}\:
\biggl[\frac{1}{n}\sum_{i=0}^{n-1}\biggl(
\frac{1}{\tau}\int_{(i+n)\tau_0}^{(i+2n)\tau_0}y(t)\,dt - 
\frac{1}{\tau}\int_{i\tau_0}^{(i+n)\tau_0}y(t)\,dt
\biggr)\biggr]^2\right\}\\
&\hspace*{0.8\textwidth}\text{MVAR}\\
&\text{with $\tau=n\tau_0$}~.
\end{split}
\label{eqn:mvar-def}
\end{align}
The above is similar to a wavelet variance
\begin{align}
\mathrm{mod}\,\sigma^2_y(\tau)
&=\mathbb{E}\left\{\Bigl[\int_{-\infty}^{+\infty} y(t)\,w_M(t) \,dt \Bigr]^2\right\}~,		
\end{align}
in which the  weight function for $\tau_0\ll\tau$, or equivalently
for $n\gg1$, can be written as
\begin{align}
w_M&=\begin{cases}
     -\frac{1}{\sqrt{2}\tau^2}t 			& 0<t<\tau\\
     \frac{1}{\sqrt{2}\tau^2}(2t-3)		& \tau<t<2\tau\\
     -\frac{1}{\sqrt{2}\tau^2}(t-3\bigr)	& 2\tau<t<3\tau\\
     0 & \text{elsewhere}
  \end{cases}
  \qquad\text{(Fig.~\ref{fig:wavelets})}~.
\end{align}
Once again, the weight function differs from a wavelet in that it is normalized for power instead of energy
\begin{align}
\mathcal{E}\{w_M\}&=\int_{-\infty}^{+\infty} w^2_M(t)\,dt = \frac{1}{2\tau}~.
\end{align}
Interestingly, it holds that $\mathcal{E}\{w_M\}=\frac12\mathcal{E}\{w_A\}$.  This is related to the fact that the AVAR response to white frequency noise $S_y(f)=h_0$ is $\smash{\sigma^2_y(\tau)=\frac{h_0}{2\tau}}$, while the response of MVAR to
the same noise is $\smash{\mathrm{mod}\,\sigma^2_y(\tau)=\frac{h_0}{4\tau}}$. 

\section{High-resolution frequency counters}\label{sec:counters}
In this section the phase noise $\phi(t)$, and therefore the frequency noise $y(t)$, is provisionally assumed to be a zero-mean stationary process. This hypothesis will no longer necessary for the direct measurement of the Allan variance.    

\subsection{Classical reciprocal counters}\label{ssec:simple-counter}
\begin{figure}[t]
\centering\includegraphics[scale=0.8]{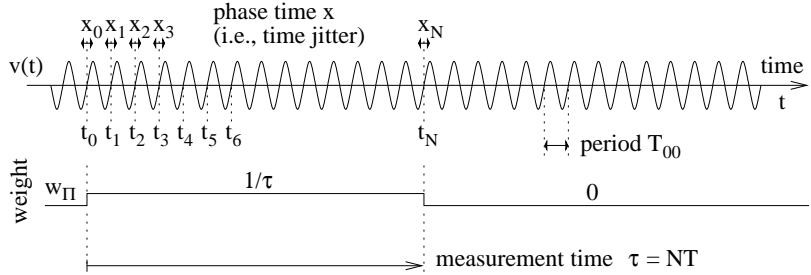}
\caption{Rectangular averaging mechanism in simple frequency counters.}
\label{fig:bare-mean}
\end{figure}
Traditionally, the uniform average over a suitable time interval $\tau$ is used as an estimator of the frequency $\nu$.  The expectation of $\nu$ is therefore
\begin{align}
&\mathbb{E}\{\nu\}=\int_{-\infty}^{+\infty} \nu(t) w_\Pi(t) \,dt && \text{$\Pi$ estimator}
\label{eqn:pi-expectation}
\\[0.5ex]
&w_\Pi(t)=\begin{cases}1/\tau & 0<t<\tau\\0 & \text{elsewhere}\end{cases}
\label{eqn:pi-weight}\\[0.5ex]
&\int_{-\infty}^{+\infty} w_\Pi(t) \, dt=1 &&\text{normalization}
~.
\end{align}
Inside, the counter measures the time interval $\tau=t_N-t_0$ between two zero crossings of $v(t)$ spaced by $N$ periods (Fig.~\ref{fig:bare-mean}).  Thus $\mathbb{E}\{\nu\}=\overline{\nu}=N/\tau$.  The averaging time $\tau_U$ selected by the user is rounded to $\tau=NT\ge\tau_U$ by stopping the measurement at the first zero crossing after that $\tau_U$ has elapsed. 
A variety of interpolation techniques~\cite{kalistz04metro} can be exploited to avoid of the uncertainty $\smash{\frac{1}{\tau\nu_c}}$ that results from the bare count of clock pulses at the frequency $\nu_c$.  The measurement of $\tau$ is affected by the error $x_0-x_N$ that results from the trigger noise and from the clock interpolator.  Here, the reference clock is assumed ideal.  Thus it holds that $\smash{y=\frac{x_N-x_0}{\tau}}$.  
With a state-of-the-art counter, the resolution of the interpolator in a single-event time interval measurement can be of $10^{-11}$~s.  
Let us assume that $x_0$ and $x_N$ are independent and have identical statistical properties, and denote with $\sigma_x^2$ the variance of each.  Under this assumption, which will be justified afterwards under more stringent conditions, the variance of $\tau$ is $2\sigma^2_x$.
Accordingly, the variance of the fractional frequency fluctuation is 
\begin{equation}
\sigma^2_y=\frac{2\sigma^2_x}{\tau^2}\qquad
\begin{array}{ll}\text{classical}\\[-0.5ex]\text{variance}\end{array}~.
\label{eqn:bare-variance}
\end{equation}
The law $\sigma^2_y\propto1/\tau^2$ is a property of the $\Pi$ estimator, i.e., of the uniform average, in the presence of white phase noise.
 
\subsection{Enhanced-resolution counters}\label{ssec:enhanced-counter}
\begin{figure}[t]
\centering\includegraphics[scale=0.8]{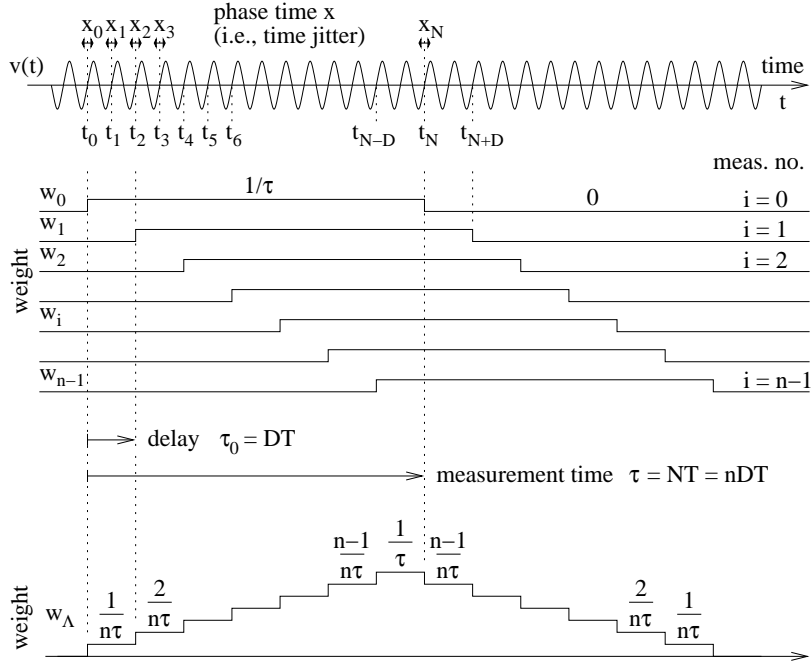}
\caption{Triangular averaging mechanism, implemented in some high-resolution frequency counters.}
\label{fig:overlap-avg}
\end{figure}
More sophisticated counters make use of the $\Lambda$ estimator (Fig.~\ref{fig:overlap-avg}), which consists of a triangular-weight average.  The counter takes a series of $n$ measures $\overline{\nu}_i=N/\tau_i$ delayed by $i\tau_0=iDT$, where 
$\tau_i=t_{N+iD}-t_{iD}$, $i\in\{0,\cdots,n-1\}$ is the time interval measured from the $(iD)$-th
to the $(N+iD)$-th zero crossings.  The expectation of $\nu$ is evaluated as the average
\begin{equation}
\mathbb{E}\{\nu\}=\frac{1}{n}\sum_{i=0}^{n-1}\overline{\nu}_i \qquad \text{where~$\overline{\nu}_i=N/\tau_i$}~.
\label{eqn:lambda-sum}
\end{equation} 
Eq.~\req{eqn:lambda-sum} can be written as an integral similar to \req{eqn:pi-expectation}, but for the weight function $w_\Pi$ replaced with $w_\Lambda$
\begin{align}
&\mathbb{E}\{\nu\}=\int_{-\infty}^{+\infty} \nu(t) w_\Lambda(t) \,dt 
  &&\text{$\Lambda$ estimator}~.
\label{eqn:lambda-expectation}
\intertext{For $\tau_0\ll\tau$, $w_\Lambda$ approaches the triangular-shape function}
&w_\Lambda(t)=\begin{cases}t/\tau&0<t<\tau\\2-t/\tau&\tau<t<2\tau\\0&\text{elsewhere}\end{cases}
\label{eqn:lambda-weight}\\[0.5ex]
&\int_{-\infty}^{+\infty} w_\Lambda(t) \,dt =1 &&\text{normalization}~.
\end{align}
Nonetheless the integral \req{eqn:lambda-expectation} is evaluated as the sum \req{eqn:lambda-sum} because the time measurements take place at the zero crossings. 
The measures $\overline{\nu}_i$ are independent because the timing errors $x_k$, $k\in\{0,\cdots,n-1\}$ are independent, as explained underneath. The counter noise is due to the interpolator noise, and to the noise of the input trigger. The samples of the interpolator jitter are independent because the interpolator is restarted every time it is used.  The trigger noise spans from dc to the trigger bandwidth $B$, which is at least the maximum switching frequency of the counter.  With modern instruments, $B$ is hardly lower than 100~MHz, hence white noise is dominant.  The autocorrelation function of the trigger noise is a sharp pulse of duration $T_R\approx1/B$.  On the other hand, the delay $\tau_0$ is lower-bounded by the period $T_{00}$ of the input signal and by the conversion time of the interpolator.  The latter may take a few microseconds.  Hence in practice it holds that $\tau_0\gg T_R$, and therefore the timing errors are independent.  Accordingly, the variance of the fractional frequency fluctuation is 
\begin{equation}
\sigma^2_y=\frac{1}{n}\frac{2\sigma^2_x}{\tau^2}
\qquad\begin{array}{ll}\text{classical}\\[-0.5ex]\text{variance}\end{array}.
\label{eqn:overlap-variance}
\end{equation}
At low input frequency, there is no reason for the delay $\tau_0$ between overlapped measures $\overline{\nu}_i$ and $\overline{\nu}_{i+1}$ to be longer than $T_{00}$, i.e., one period.  Thus $D=1$, $\tau_0=T_{00}$, and $n=N=\nu_{00}\tau$.  Hence Eq.~\req{eqn:overlap-variance} is rewritten as
\begin{equation}
\sigma^2_y=\frac{1}{\nu_{00}}\frac{2\sigma^2_x}{\tau^3}
\qquad\begin{array}{ll}\text{classical}\\[-0.5ex]\text{variance}\end{array}.
\label{eqn:overlap-variance-lowf}
\end{equation}
At high  input frequency, the minimum delay $\tau_0$ is set by the conversion time of the interpolator.  Hence the measurement rate is limited to $\nu_I$ measures per second, the number $n$ of overlapped measures is $n=\nu_I\tau\le\nu_{00}\tau$, and Eq.~\req{eqn:overlap-variance} turns into
\begin{equation}
\sigma^2_y=\frac{1}{\nu_I}\frac{2\sigma^2_x}{\tau^3}
\qquad\begin{array}{ll}\text{classical}\\[-0.5ex]\text{variance}\end{array}.
\label{eqn:overlap-variance-highf}
\end{equation}
The law $\sigma^2\propto1/\tau^3$, either  \req{eqn:overlap-variance-lowf} or \req{eqn:overlap-variance-highf}, is a property of the $\Lambda$ estimator in the presence of white noise.  This property is the main reason for having introduced the $\Lambda$ estimator in frequency counters.  Yet it is to be made clear that the enhanced resolution is achieved by averaging on multiple measurements, even though overlapped, and that the measurement of a single event, like a start-stop time interval, can not be improved in this way.

\subsection{Understanding technical information}
Searching through the instruction manual of frequency counters and through the manufacturer web sites, one observes that the problem of the estimation is generally not addressed.
When the counter is of the $\Pi$ type, the measurement mechanism is often explained with a figure similar to Fig.~\ref{fig:bare-mean}.  On the other hand, the explanation for the overlapped measurements in $\Lambda$-type counters is not found in the technical documentation.  
As a further element of confusion, both counters provide one value every $\tau$ seconds when programmed to measure over the time $\tau$. This can lead the experimentalist to erroneously assume that the estimation is always of the $\Pi$ type. 

The internal estimation mechanism can be understood from the formula for the ``frequency error'', often given in the technical documentation.  These formulae are of the form   
\begin{align}
(\Pi)\quad\sigma_y&=\frac{1}{\tau}\, \sqrt{2(\delta t)^2_\mathrm{trigger} + 
	2(\delta t)^2_\mathrm{interpolator}}
\label{eqn:bare-variance-spec}
\intertext{or}
\begin{split}
(\Lambda)\quad\sigma_y&=\frac{1}{\tau\sqrt{n}}\, \sqrt{2(\delta t)^2_\mathrm{trigger} + 
	2(\delta t)^2_\mathrm{interpolator}}\\[0ex]
&n=\begin{cases}
     \nu_0\tau  &\nu_{00}\le\nu_I\\
     \nu_I\tau  &\nu_{00}>\nu_I
  \end{cases}
\end{split}
\label{eqn:overlap-variance-spec}
\end{align}
where $\nu_I$ is of the order of 200~kHz.
The actual formulae may differ in that uncertainty and noise of the reference frequency may be included or not; in that the factor 2 in the interpolator noise does not appear explicitely; and in other details.  

The terms inside the square root of \req{eqn:bare-variance-spec} and \req{eqn:overlap-variance-spec} come from independent white noise processes, as explained in Sections \ref{ssec:simple-counter}. Thus, one can match Eq.~\req{eqn:bare-variance-spec} to \req{eqn:bare-variance}, and Eq.~\req{eqn:overlap-variance-spec} to \req{eqn:overlap-variance}.  Consequently, the presence of a term $\tau$ in the denominator reveals that the counter is of the $\Pi$ type, while the presence of the term $\smash{\tau\sqrt{n}}$ or $\smash{\tau\sqrt{\tau}}$ reveals that the counter is of the $\Lambda$ type.

\subsection{Examples}
Two instruments have been selected with the sole criteria that the author is familiar
with them, and that they are well suitable to show how to the $\Pi$ and the $\Lambda$
estimators can be identified in the instruction manual.   

\paragraph{Stanford Research Systems SR-620 ($\Pi$ estimator).}  In the 
instruction manual  \cite[p.\,27]{stanford:srs620-counter}, the RMS uncertainty
is called resolution and given by the formula
\begin{align}
\left[\!\!\!\begin{array}{c}\text{\footnotesize RMS}\\[-0.7ex]
   \text{\footnotesize resolution}\\[-0.7ex]
   \text{\footnotesize (in Hz)}\end{array}\!\!\!\right]=
\frac{\text{\footnotesize frequency}}{\text{\footnotesize gate time}} 
\sqrt{\frac{(25\:\mathrm{ps})^2
+\left[\left(\!\!\!\begin{array}{c}\text{\footnotesize short term}\\[-1ex]
   \text{\footnotesize stability}\end{array}\!\!\!\right)
\times\left(\!\!\!\begin{array}{c}\text{\footnotesize gate}\\[-1ex]
  \text{\footnotesize time}\end{array}\!\!\!\right)\right]^2
+2{\times}\left[\!\!\!\begin{array}{c}\text{\footnotesize trigger}\\[-1ex]
   \text{\footnotesize jitter}\end{array}\!\!\!\right]^2}{\mathrm{N}}}~.
\label{eqn:resolution-stanford}
\end{align}
The above \req{eqn:resolution-stanford} matches our notation with
\begin{center}\begin{tabular}{ll}
RMS resolution & $\sigma_\nu=\nu_{00}\sigma_y$\quad(classical variance)\\
frequency         & $\nu_{00}$\\
gate time          & $\tau$
\end{tabular}\end{center}
The numerator inside the square root is the square single-shot time deviation $2\sigma^2_x$.  
It includes the inherent resolution of the counter, 25 ps, ascribed to the interpolator; 
the phase noise of the frequency standard; and the equivalent input noise 
(signal plus trigger) divided by the input slew rate.  
The denominator ``$\mathrm{N}$'' is the number 
of measurement averaged, not to be mistaken for $N$ of our notation.
If the measurement time is equal to the gate time, it holds that $\mathrm{N}=1$.  
Eq.\ \req{eqn:resolution-stanford} divided by $\nu_{00}$ and squared
is of the same form of Eq.\ \req{eqn:bare-variance}, thus of 
Eq.\ \req{eqn:bare-variance-spec}, with $\sigma^2_y\propto\smash{\frac{1}{\tau^2}}$.
This indicates that the counter is of the $\Pi$ type.

\paragraph{Agilent Technologies 53132A  ($\Lambda$ estimator).}
The instruction manual \cite[pp.\ 3-5 to 3-8]{agilent:53131-counter} reports the
RMS resolution given by the formula
\begin{align}
&\left[\!\!\!\begin{array}{c}\text{\footnotesize RMS}\\[-0.7ex]
   \text{\footnotesize resolution}\end{array}\!\!\!\right]=
\left(\!\!\!\begin{array}{c}\text{\footnotesize frequency}\\[-1ex]
  \text{\footnotesize or period}\end{array}\!\!\!\right)
\times \left[
\frac{4{\times}\sqrt{(t_\mathrm{res})^2+2\times(\text{\footnotesize trigger error})^2}}{%
  (\text{\footnotesize gate time})\times\sqrt{\text{\footnotesize no.\ of samples}}}
+\frac{t_\mathrm{jitter}}{\text{\footnotesize gate time}}
\right]
\label{eqn:resolution-agilent}\\
&\begin{array}{ll}
t_\mathrm{res}=225~\mathrm{ps}&\\
t_\mathrm{jitter}=3~\mathrm{ps}&\\[1ex]
\text{number of samples}=\begin{cases}
(\text{gate time})\times(\text{frequency})&\text{for}~f<200\:\text{kHz}\\
(\text{gate time})\times2{\times}10^5&\text{for}~f\ge200\:\text{kHz}
\end{cases}
\end{array}\nonumber
\end{align}
The above \req{eqn:resolution-agilent} matches our notation with
\begin{center}\begin{tabular}{ll}
RMS resolution & $\sigma_\nu=\nu_{00}\sigma_y$ or 
                            $\sigma_T=T_{00}\sigma_y$\quad(classical variance)\\
frequency         & $\nu_{00}$\\
period               & $T_{00}$\\
gate time          & $\tau$\\
no.\ of samples & $n$
\end{tabular}\end{center}
The term $\smash{\frac{t_\mathrm{jitter}}{\text{gate time}}}$ 
of \req{eqn:resolution-agilent} is due to some internal phenomena not analyzed
here, for we provisionally neglect it.
The numerator inside the square brackets is the 
single-shot time deviation $\smash{2\sigma^2_x}$.  After squaring and normalizing, 
Eq.\ \req{eqn:resolution-agilent} is of the same form of \req{eqn:overlap-variance}, thus of \req{eqn:overlap-variance-spec}.
The counter switches averaging mode at the input frequency of $\nu_I=200$ kHz.
In fact, it holds that
\begin{align}
n=\begin{cases}
\nu_{00}\tau& \nu_{00}<200~\mathrm{kHz}\\
\tau\times2{\times}10^5& \nu_{00}\ge200~\mathrm{kHz}
\end{cases}\nonumber
\end{align}
Accordingly, Eq.\ \req{eqn:resolution-agilent} turns into \req{eqn:overlap-variance-lowf}
for $\nu<2{\times}10^5$~Hz, and into \req{eqn:overlap-variance-highf} beyond
$2{\times}10^5$~Hz. The variance is $\sigma^2_y\propto\smash{\frac{1}{\tau^3}}$
in both cases, for the counter is of the $\Lambda$ type.

\paragraph{Remarks.} The resolution enhancement of the $\Lambda$ average
can only be used with stationary phenomena, not with single events.  This is easily seen in the table underneath.  This table compares the two counters, SR-620 and 53132A,
under the simplified assumptions that the reference clock is ideal, and that the high slew rate makes the trigger noise negligible.
\begin{center}\begin{tabular}{ccc}
counter  & single-event start-stop & frequency $\nu_{00}=100$\:kHz\\[-0.3ex]
type	      & time interval $\mathrm{TI}=1$ s & averaged on $\tau=1$\:s\\\hline 
SR-620  & $\sigma=2.5{\times}10^{-11}$ & $\sigma_y=2.5{\times}10^{-11}$
\rule[0ex]{0ex}{2.5ex}\\
53132A  & $\sigma=9{\times}10^{-10}$ & $\sigma_y=5.8{\times}10^{-12}$
\end{tabular}\end{center} 
The superior resolution of the SR620 interpolator provides higher 
resolution for the measurement of a single event in start-stop mode.
On the other hand, the highest resolution is obtained with the
53132A in the measurement of stationary signals, due to the $\Lambda$ 
estimator.

\section{Frequency stability measurement}\label{sec:avar-meas}
\begin{figure}[t]
\centering\includegraphics[scale=0.8]{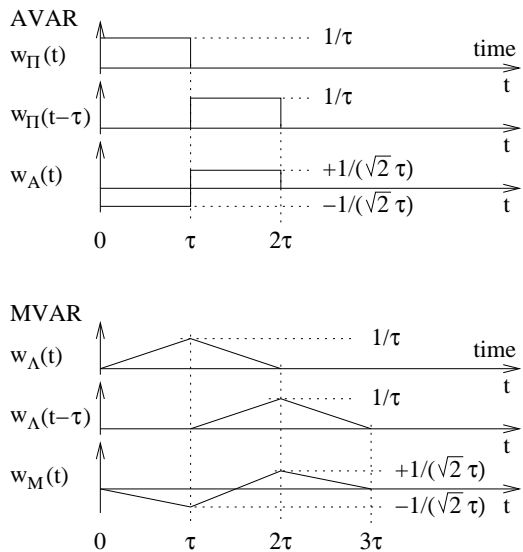}
\caption{Relationships between the weight functions.}
\label{fig:wavelet-build}
\end{figure}
Let us first observe that it holds
\begin{align}
w_A(t)&=\frac{1}{\sqrt2}\Bigl[w_\Pi(t-\tau) - w_\Pi(t)\Bigr]
&\text{(Fig.\ \ref{fig:wavelet-build})}\\
w_M(t)&=\frac{1}{\sqrt2}\Bigl[w_\Lambda(t-\tau) - w_\Lambda(t)\Bigr]
&\text{(Fig.\ \ref{fig:wavelet-build})}
&~.
\label{eqn:wlambda-to-wm}
\end{align}

\subsection{$\Pi$-type counters}
\paragraph{AVAR\@.}
Let us get a stream of data $\smash{\overline{y}_k^{(1)}}$ from the output of a $\Pi$-estimator counter,  measured over a base time slot $\tau_B$ with zero dead time.
The superscript ``$^{(1)}$'' refers to the averaging time $1\times\tau_B$.
Feeding this stream into Eq.\ \req{eqn:avar-def}, one gets Eq.\ \req{eqn:avar-expanded} with $\tau=\tau_B$.  Then, after averaging contiguous data in groups of $m$, one gets a smaller file of $\smash{\overline{y}_k^{(m)}}$, averaged over the time $m\tau_B$. Feeding this new file into Eq.\ \req{eqn:avar-def}, one gets Eq.\ \req{eqn:avar-expanded} with $\tau=m\tau_B$. This is exactly what one expects.  It is a common practice to
plot of $\smash{\sigma^2_y(\tau)}$ in this way, using a single data stream
and $m$ in powers of 2.

\paragraph{MVAR\@.}
The data file $\smash{\overline{y}_k^{(1)}}$ can also be fed in Eq.\ \req{eqn:mvar-def}.  In this case the measurement time $\tau_B$ of the counter is the delay $\tau_0$ of \req{eqn:mvar-def}.  The variance is evaluated at $\tau=n\tau_B$ set by the choice
of $n$.   Yet, it is desirable that $n\gg1$.

\subsection{$\Lambda$-type counters.}
The attention of the reader is now to be drawn to a subtilty in the use of a $\Lambda$-estimator to measure the Allan variance.  While the counter provides one value of $\overline{\nu}$ every $\tau_B$ seconds, two contiguous windows $w_\Lambda(t)$ and $w_\Lambda(t-\tau_B)$ are overlapped by $\tau_B$.  That is, the falling side of $w_\Lambda(t)$ overlaps the rising side of $w_\Lambda(t-\tau_B)$.  Thus $\overline{\nu}_{k+1}-\overline{\nu}_k$, hence 
$\overline{y}_{k+1}-\overline{y}_k$, is the frequency averaged with the $w_M$ window.  
The practical consequence is that, feeding the file of such $\smash{\overline{y}_k^{(1)}}$ into Eq.\ \req{eqn:avar-def} (AVAR), one gets \emph{exactly} Eq.\ \req{eqn:mvar-def} with $\tau=\tau_B$.  That is, MVAR instead of AVAR.

\begin{figure}[t]
\centering\includegraphics[scale=0.8]{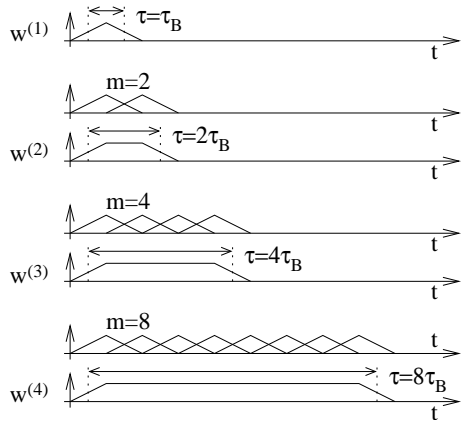}
\caption{The average of $m$ contiguous $\Lambda$ measures converges to a $\Pi$ estimate.}
\label{fig:joined-slots}
\end{figure}

A longer measurement time $\tau$ is obtained by averaging contiguous data in groups of $m$.  This process yields a smaller file of $\smash{\overline{y}_k^{(m)}}$, averaged over $m\tau_B$. Yet the measurements are overlapped, for the weight function, which is the triangle $w_\Lambda$ for $m=1$, turns into an isosceles trapezium for $m>1$, and asymptotically into the rectangle $w_\Pi$ for large $m$. 
A graphical proof is proof is given in Fig.\ \ref{fig:joined-slots}. 
The practical consequence is that, feeding the file of $\smash{\overline{y}_k^{(m)}}$ into Eq.\ \req{eqn:avar-def}, one gets an ``odd'' variance for with small $m>1$, i.e., a variance that is neither AVAR or MVAR; and again the AVAR for large $m$.  

Unfortunately, the conversion between $\Pi$ and $\Lambda$ estimates is impossible
without combining a large number of data.  It is therefore impossible to measure the AVAR with a $\Lambda$-type counter without increasing $\tau$, that is, $\tau=m\tau_B$ with a sufficiently large value of $m$.

Finally, we focus the attention to two asymptotic classes of measurement, namely short-term stability and long-term stability, analyzed underneath.

\paragraph{Long-term stability.}  These measurements are characterized by a large value of $m$. Hence the averaging function converges to $w_\Pi$, thus the data file $\smash{\overline{y}_k^{(m)}}$ can be fed into the AVAR formula \req{eqn:avar-def} without mistakes.  This is a fortunate outcome for two reasons, namely 
\begin{enumerate}
\item AVAR is preferable to MVAR because, for a given duration of the time record, it
provides a more accurate estimate at large $\tau$,
\item the rejection of white phase noise of the MVAR is not necessary in long term measurements.
\end{enumerate}

\paragraph{Short-term stability.}   The experimentalist interested in short-term stability appreciates the rejection of white noise of the MVAR, while the longer duration of the experiment, as compared to the measurement of the AVAR at the same $\tau$ is not disturbing. In this case, the bare mean can not be used to combine contiguous values in order to get $\tau=m\tau_B$.  The values must be weighted proportionally to the triangular staircase sequence $\{1, 2, \ldots, \lceil m/2\rceil, \ldots\,, 2, 1\}$, so that the equivalent weight function is an isosceles triangle of width $2\tau$.  A file of such measures fed into \req{eqn:avar-def} gives the MVAR evaluated at $\tau=m\tau_B$. The nuisance is that this triangular-shape average is only possible for odd $m$.

\subsection{Other methods}
For long-term measurements, the \emph{total variance} TotVar \cite{greenhall99uffc} is progressively being used as an estimator of the AVAR\@.  The importance of the TotVar resides in that, given a record if measured data, it exhibits higher accuracy at larger $\tau$. Yet TotVar is based on rectangular averages.  Consequently, the raw readout of a $\Lambda$ estimator can not be fed in the formula for TotVar without interpretation mistakes.

Another useful method, often called \emph{picket-fence},
consists of the absolute timing of the zero crossings versus a free-running time scale.  
In this case, there is no preprocessing inside the counter, for there can be no ambiguity in the result interpretation.  Absolute-time data can be used to calculate AVAR, MVAR, TotVar and other variances.
The concept of picket-fence was first proposed in \cite[Eq.\,(79)]{barnes71im}.  Then,
it was studied extensively as an independent method for the measurement of AVAR and MVAR \cite{greenhall89uffc,greenhall97im}.

\subsection*{Acknowledgments}
I wish to thank Vincent Giordano (FEMTO-ST, Besan\c{c}on, France) for
pointing out the need for this work; John Dick, Charles Greenhall (JPL, Pasadena, CA) and David Howe (NIST, Boulder, CO) for stimulating discussions.  I wish to thank
Fran\c{c}ois Vernotte (Observatoire de Besan\c{c}on, France) and Mark Oxborrow 
(NPL, Teddington, UK) for discussions and thorough revision of the early manuscript
versions.

\def\bibfile#1{/Users/rubiola/Documents/work/bib/#1}
\bibliographystyle{amsalpha}
\bibliography{\bibfile{ref-short},%
              \bibfile{references},%
              \bibfile{rubiola}}

\newcommand{\etalchar}[1]{$^{#1}$}
\providecommand{\bysame}{\leavevmode\hbox to3em{\hrulefill}\thinspace}
\providecommand{\MR}{\relax\ifhmode\unskip\space\fi MR }
\providecommand{\MRhref}[2]{%
  \href{http://www.ams.org/mathscinet-getitem?mr=#1}{#2}
}
\providecommand{\href}[2]{#2}
\begin{thebibliography}{BCC{\etalchar{+}}72}

\bibitem[AB81]{allan81fcs}
David~W. Allan and James~A. Barnes, \emph{A modified ``{A}llan variance'' with
  increased oscillator characterization ability}, Proc. 35 FCS (Ft.~Monmouth,
  NJ), May 1981, pp.~470--474.

\bibitem[Agi98]{agilent:53131-counter}
Agilent Technologies, Inc., Paloalto, CA, \emph{{53131A} and {53132A}
  instruction manual (part number 53131-90055)}, 1998.

\bibitem[All66]{allan66pieee}
David~W. Allan, \emph{Statistics of atomic frequency standards}, Proc.\ IEEE
  \textbf{54} (1966), no.~2, 221--230.

\bibitem[BCC{\etalchar{+}}72]{barnes71im}
James~A. Barnes, Andrew~R. Chi, Leonard~S. Cutler, Daniel~J. Healey, David~B.
  Leeson, Thomas~E. Mc{G}unigal, James~A. Mullen, Jr, Warren~L. Smith,
  Richard~L. Sydnor, Robert F.~C. Vessot, and Gernot M.~R. Winkler,
  \emph{Characterization of frequency stability}, IEEE Trans.\ Instrum.\ Meas.
  \textbf{20} (1972), 105--120.

\bibitem[{CCI}90]{ccir90rep580-3}
{CCIR Study Group VII}, \emph{Characterization of frequency and phase noise,
  {R}eport no.\ 580-3}, Standard Frequencies and Time Signals, Recommendations
  and Reports of the {CCIR}, vol. {VII} (annex), International
  Telecommunication Union {(ITU)}, Geneva, Switzerland, 1990, pp.~160--171.

\bibitem[GHP99]{greenhall99uffc}
Charles~A. Greenhall, Dave~A. Howe, and Donald~B. Percival, \emph{Total
  variance, an estimator of long-term frequency stability}, IEEE Trans.\
  Ultras.\ Ferroel.\ and Freq.\ Contr. \textbf{46} (1999), no.~5, 1183--1191.

\bibitem[Gre89]{greenhall89uffc}
Charles~A. Greenhall, \emph{A method for using time interval counters to
  measure frequency stability}, IEEE Trans.\ Ultras.\ Ferroel.\ and Freq.\
  Contr. \textbf{36} (1989), no.~5, 478--480.

\bibitem[Gre97]{greenhall97im}
\bysame, \emph{The third-difference approach to modified allan variance}, IEEE
  Trans.\ Instrum.\ Meas. \textbf{46} (1997), no.~3, 696--703.

\bibitem[Kal04]{kalistz04metro}
J\'ozef Kalisz, \emph{Review of methods for time interval measurements with
  picosecond resolution}, Metrologia \textbf{41} (2004), 17--32.

\bibitem[Kro83]{kroupa:frequency-stability}
Ven{\v{c}}eslav~F. Kroupa (ed.), \emph{Frequency stability: Fundamentals and
  measurement}, IEEE Press, New York, 1983.

\bibitem[LA84]{lesage84im}
Paul Lesage and Th\'eophane Ayi, \emph{Characterization of frequency stability:
  Analysis of the modified {A}llan variance and properties of its estimate},
  IEEE Trans.\ Instrum.\ Meas. \textbf{33} (1984), no.~4, 332--336.

\bibitem[Rut78]{rutman78pieee}
Jacques Rutman, \emph{Characterization of phase and frequency instabilities in
  precision frequency sources: Fifteen years of progress}, Proc.\ IEEE
  \textbf{66} (1978), no.~9, 1048--1075.

\bibitem[Sny80]{snyder80ao}
J.~J. Snyder, \emph{Algorithm for fast digital analysis of interference
  fringes}, Appl.\ Opt. \textbf{19} (1980), no.~4, 1223--1225.

\bibitem[Sta04]{stanford:srs620-counter}
Stanford Research Systems, Inc., \emph{{SR620} universal time interval counter
  instruction manual rev.~2.6}, 2004.

\bibitem[{Vig}99]{ieee99std1139}
John~R. {Vig (chair.)}, \emph{{IEEE} standard definitions of physical
  quantities for fundamental frequency and time metrology--random instabilities
  ({IEEE} standard 1139-1999)}, {IEEE}, New York, 1999.

\bibitem[VZL98a]{vernotte98metro-b}
F.~Vernotte, G.~Zalamansky, and E.~Lantz, \emph{Time stability characterization
  and spectral aliasing. {P}art {II}: a frequency-domain approach}, Metrologia
  \textbf{35} (1998), 731--738.

\bibitem[VZL98b]{vernotte98metro-a}
\bysame, \emph{Time stability characterization and spectral aliasing. {P}art
  {II}: a time-domain approach}, Metrologia \textbf{35} (1998), 723--730.

\end{thebibliography}

\end{document}